\documentclass[twocolumn,twocolappendix]{aastex631}
\usepackage{graphicx}
\usepackage[T1]{fontenc}
\usepackage[utf8]{inputenc}
\usepackage{bm}
\usepackage{amsmath}

\newcommand{\Add}[1]{\textcolor{black}{}}

\newcommand{\pfrac}[2]{ \biggl(\dfrac{#1}{#2}\biggr) }

\received{January 8, 2025}
\revised{February 27, 2025}
\accepted{March 3, 2025}

\begin{document}

\title{Investigating the Bouncing Barrier with Collision Simulations of Compressed Dust Aggregates}

\author[0009-0004-3868-373X]{Haruto Oshiro}
\email{oshiro.h.aa@m.titech.ac.jp}
\affiliation{Department of Earth and Planetary Science, Institute of Science Tokyo, 2-12-1, Ookayama, Megro, Tokyo 152-8550, Japan}

\author[0000-0003-1844-5107]{Misako Tatsuuma}
\affiliation{RIKEN Interdisciplinary Theoretical and Mathematical Sciences Program (iTHEMS), 2-1 Hirosawa, Wako, Saitama 351-0198, Japan}
\affiliation{Department of Earth and Planetary Science, Institute of Science Tokyo, 2-12-1, Ookayama, Megro, Tokyo 152-8550, Japan}

\author[0000-0002-1886-0880]{Satoshi Okuzumi}
\affiliation{Department of Earth and Planetary Science, Institute of Science Tokyo, 2-12-1, Ookayama, Megro, Tokyo 152-8550, Japan}

\author[0000-0001-9659-658X]{Hidekazu Tanaka}
\affiliation{Astronomical Institute, Graduate School of Science, Tohoku University, 6-3 Aramaki, Aoba-ku, Sendai 980-8578, Japan}



\begin{abstract}

The collision outcomes of dust aggregates in protoplanetary disks dictate how planetesimals form.
Experimental and numerical studies have suggested that bouncing collisions occurring at low impact velocities may limit aggregate growth in the disks, but the conditions under which bouncing occurs have yet to be fully understood.
In this study, we perform a suite of collision simulations of moderately compact dust aggregates with various impact velocities, aggregate radii, and filling factors ranging between 0.4 and 0.5. 
Unlike previous simulations, we generate compact aggregates by compressing more porous ones, mimicking the natural processes through which compact aggregates form.
We find that the compressed aggregates bounce above a threshold mass, which decreases with impact velocity. 
The threshold mass scales with impact velocity as the $-4/3$ power, consistent with the findings of previous experiments. 
We also find that the threshold aggregate mass for bouncing depends strongly on filling factor, likely reflecting the strong filling-factor dependence of the compressive strength of compressed aggregates.
Our energy analysis reveals that nearly 90\% of the initial impact energy is dissipated during the initial compression phase, and over 70\% of the remaining energy is dissipated during the subsequent stretching phase, regardless of whether the collision results in sticking or bouncing.
Our results indicate that dust aggregates with a filling factor of 0.4 cease to grow beyond 100 $\mathrm{\mu m}$ as a result of the bouncing barrier.

\end{abstract}

\keywords{Planet formation (1117); Planetesimals (1119); Protoplanetary disk (1118); Collisional process (1160); Dust physics (3087)}


\section{Introduction} \label{sec:intro}
Coagulation of submicron-sized dust grains in protoplanetary disks is the first step in planet formation. 
Submicron-sized dust grains collide and form dust aggregates, which lead to the formation of planetesimals through collisions \citep[e.g.,][]{Okuzumi+12,Kataoka+13b,Kobayashi&Tanaka21} and/or some instabilities \citep[e.g.,][]{Goldreich&Ward73,Youdin&Goodman05,Johansen+07,Bai&Stone10,Takahashi+14,Tominaga+21}. 
Understanding the growth path of dust aggregates is important as their physical properties, such as size, internal structure, and material, affect the thermal structure and observational appearance of protoplanetary disks \citep[e.g.,][]{Tanaka+05,Johansen+09,Testi+14,Kataoka+15,Tazaki+19,Lesur+23}.

Bouncing, one of the collisional outcomes of compact aggregates, is a mechanism that can inhibit dust growth in protoplanetary disks {\citep{Guttler+10,Zsom+10}}. 
Previous simulations suggest that bouncing has a significant effect on dust growth in protoplanetary disks \citep[e.g.,][]{Zsom+10,Windmark+12,Stammler+23,Dominik&Dullumond24}. 
Millimeter-wave polarimetric observations of protoplanetary disks point to the existence of relatively compact aggregates \citep[e.g.,][]{Tazaki+19,Zhang+23}, which may have formed by bouncing collisions {\citep{Weidling+09,Dominik&Dullumond24}}.
Understanding the bouncing barrier is thus important for the growth limits of dust aggregates in protoplanetary disks.

The conditions under which bouncing occurs have been studied through laboratory experiments and numerical simulations \citep[e.g.,][]{Langkowski+08,Guttler+10,Wada+11,Weidling+12,Schrapler+12,Seizinger+Kley13,Kothe+13,Schrapler+22,Arakawa+23}, but there are some discrepancies between them. 
Experiments typically show that aggregates with filling factors above 0.15 bounce with a certain range of impact velocities \citep{Langkowski+08}, and that larger aggregates tend to bounce at lower velocities \citep[e.g.,][]{Guttler+10,Weidling+12,Schrapler+22}. 
In particular, \citet{Kothe+13} studied the outcome of free collisions between sub-mm-sized dust aggregates consisting of micron-sized $\mathrm{SiO_2}$ spheres, 
finding that the threshold aggregate mass for collisional sticking/bouncing is proportional to the $-4/3$ power of impact velocity.
However, these experimental results are not fully consistent with the results of previous numerical simulations \citep{Wada+11,Seizinger+Kley13,Arakawa+23}.
Simulations by \citet{Wada+11} and \citet{Seizinger+Kley13} only confirmed bouncing of aggregates that are considerably less porous (i.e., filling factor $\gtrsim0.4$) than those used in experiments.
More recent simulations by \citet{Arakawa+23} 
have shown that larger aggregates are more likely to bounce, but their results still do not show a dependence of the bouncing threshold mass on impact velocity.

In this study, we explore the possibility that the discrepancies in the previous experiments and simulations are due to the way the dust aggregates are formed.
Simulations by \citet{Arakawa+23} used closest-packing-and-particle-extraction (CPE) aggregates, which are made by randomly extracting particles from the closest-packed aggregates. 
In contrast, experiments by \citet{Kothe+13} used aggregates that naturally formed in dust-storage containers. 
It is possible that the artificially formed CPE aggregates and naturally formed aggregates have considerably different internal structures.
The aggregates used in the experiments have moderately high filling factors of $\approx 0.3$--0.4, suggesting that they would have experienced compaction in the containers.
Compressed aggregates are likely to better represent compact dust aggregates observed in protoplanetary disks, which are thought to have been originally fluffy but later experienced compaction \citep[e.g.,][]{Ormel+07,Okuzumi+09,Okuzumi+12,Kataoka+13b}. 
To understand the bouncing behavior of naturally forming dust aggregates, we perform collision simulations of compressed aggregates with various impact velocities, aggregate sizes, and aggregate volume filling factors. 
We show that the threshold aggregate mass for bouncing increases as the impact velocity decreases, consistent with previous experiments, and drops sharply with increasing aggregate filling factor.

This paper is structured as follows.
In Section~\ref{sec:models}, we describe our numerical model based on \citet{DT97} and \citet{Wada+07} and details of the initial aggregates. 
In Section~\ref{sec:result}, we present the results of our numerical simulations.
The energetics of aggregate collisions is analyzed  in Section~\ref{sec:energy}. 
In Section~\ref{sec:discussion}, we present our interpretation of the simulation results, compare our results with those of previous studies, and discuss implications for planet formation and limitations of our simulations. 
We summarize our results in Section~\ref{sec:conclusion}.

\section{Simulation Models} \label{sec:models}
We perform three-dimensional $N$-body simulations of aggregate collisions. 
Section~\ref{sec:interaction_model} describes the grain interaction model 
and parameter sets adopted in our simulations. Section~\ref{sec:initial_agg} introduces the method used to prepare our initial aggregates and highlights their key properties. Section~\ref{sec:col_sim} describes the parameter values adopted in our collision simulations.

\subsection{Monomer Interaction Model} \label{sec:interaction_model}
We consider aggregates made of equal-sized spherical grains, which we call monomers, and
calculate the interaction of monomers in contact using the model of \citet{DT97} and \citet{Wada+07}. 
The model is based on the Johnson--Kendall--Roberts (JKR) theory \citep{JKR+71} for the normal force between elastic spheres with adhesive forces, but also takes into account tangential forces arising from sliding, rolling, and twisting displacements. 
A detailed description of the model can be found in \citet{Wada+07}. 
Here, we only introduce the key features and parameters of the model.

The normal force between two monomers in contact is given by the sum of the elastic (repulsive) and adhesive (attractive) forces. 
In the equilibrium state, where the elastic force is balanced by surface tension, the contact radius and the normal displacement are given by
\begin{align}
    a_0&=\left[\frac{9\pi\gamma r_1^2 (1-\nu^2)}{2\varepsilon}\right]^{1/3}, \\
    \delta_0&=\frac{2a_0^2}{3r_1},
\end{align}
where $r_1$, $\gamma$, $\nu$, and $\varepsilon$ are the monomer radius, surface energy, Poisson's ratio, and Young's modulus, respectively. Here, the normal displacement $\delta$ of two equal-sized monomers is generally defined as $2r_1$ minus the distance between the monomers' centers, with $\delta > 0$ and $<0$ indicating that the monomers are mutually compressed and stretched, respectively.
The normal force required to separate the two monomers is $F_{\mathrm{c}}=(3/2)\pi\gamma r_1$.
Monomers in contact separate if $\delta$ reaches $-\delta_{\mathrm{c}}$, where $\delta_\mathrm{c}=(9/16)^{1/3}\delta_0$. The energy $E_{\mathrm{break}}$ required to separate two contacting monomers from the equilibrium state is given by
\begin{equation}
    E_{\mathrm{break}}=\left(\frac{4}{5}6^{1/3}+\frac{4}{45}\right)F_{\mathrm{c}}\delta_{\mathrm{c}}
    \approx 1.54F_{\mathrm{c}}\delta_{\mathrm{c}}.
\end{equation}

In our simulations, the force, length, and mass are normalized by $F_{\mathrm{c}}$, $\delta_{\mathrm{c}}$, and $m_1$, respectively, where $m_1=(4\pi/3)\rho_1r_1^3$ is the monomer mass and $\rho_1$ is the material density of the individual monomers. 
The unit time is given by
\begin{equation}
    t_{\mathrm{c}}=\sqrt{\frac{m_1\delta_{\mathrm{c}}}{F_{\mathrm{c}}}}.
\end{equation}

The tangential forces are elastic for small displacements less than the critical values, but behave as friction for displacements exceeding these values.
For instance, rolling friction acts on the contact surface of two monomers if their rolling displacement exceeds the critical value $\xi_{\rm crit}$. 
Rolling one monomer over another for a distance of $(\pi/2)r_1$ against rolling friction requires an energy of
\begin{equation}
    E_{\mathrm{roll}}=6\pi^2\gamma r_1\xi_{\mathrm{crit}}.
\end{equation}

\begin{table}[t]
    \centering
    \caption{Parameters for Ice Monomers Used in the Simulations.}
    \begin{tabular}{lc} \hline \hline
     Parameters & Values \\ \hline
     Monomer radius $r_1~(\mathrm{\mu m})$ & $0.1$\\
     Material density $\rho_1~(\mathrm{g\,cm^{-3}})$ & $1.0$ \\
     Surface energy $\gamma~(\mathrm{mJ\,m^{-2}})$ & $100$ \\
     Poisson's ratio $\nu$ & $0.25$ \\
     Young's modulus $\varepsilon~(\mathrm{GPa})$ & $7$ \\
     Critical rolling displacement $\xi_{\mathrm{crit}}$ (\AA) & $8$ \\ \hline
    \end{tabular}
    \tablecomments{These values are taken from \citet{Israelachvili92} and \citet{DT97}.}
    \label{tab:monomar_parmas}
\end{table}

Table~\ref{tab:monomar_parmas} summarizes the values of the material parameters adopted in our simulations.
These values are the same as those used by \citet{Wada+11} and \citet{Arakawa+23}. 
The resulting monomer mass is $m_1 = 4.2\times10^{-15}~\mathrm{g}$. 
We consider ice particles because we are particularly interested in the outer and cold region of protoplanetary disks, where the presence of relatively compact aggregates has been inferred from millimeter-wave polarimetric observations \citep[e.g.,][]{Tazaki+19,Zhang+23,Ueda+24}. 
Since the fragmentation threshold velocity scales with material parameters \citep[e.g.,][]{Wada+13}, we expect that the bouncing condition also scales with them. 
We note that some recent experiments \citep{Gundlach+18,Musiolik&Wurm19} suggest lower surface energies for low-temperature ice than adopted here. 
The dependence of the bouncing condition on surface energy will be explored in future work.

\subsection{Preparation of Initial Aggregates} \label{sec:initial_agg}

\begin{figure*}[t]
    \centering
    \includegraphics[height=4cm]{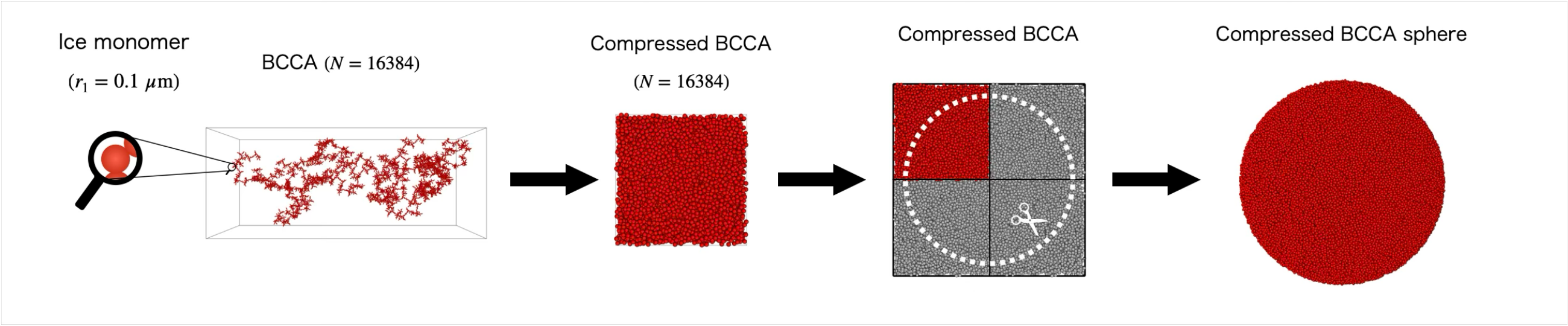}
    \caption{Schematic of how to create initial aggregates. First, we generate BCCAs consisting of 16384 icy monomers. Second, we compress each BCCA by moving periodic boundaries \citep{Tatsuuma+23}. Third, we connect identical compressed BCCAs and cut out a sphere of a given radius.}
    \label{fig:make_initial}
\end{figure*}

We create spherical aggregates with moderate volume filling factors $\phi = 0.4$--$0.5$ using the following procedure (Figure~\ref{fig:make_initial}).
First, we generate four samples of porous aggregates, each composed of $16384$ monomers, using the ballistic cluster--cluster aggregation (BCCA) algorithm \citep{Meakin91}, where aggregates are formed by making two smaller and identical aggregates along a ballistic trajectory without internal restructuring upon impact. 
The relative orientation and impact parameter of the colliding aggregates are chosen randomly, resulting in four BCCA aggregate samples with an identical mass but different internal structures. 
Aggregates formed in this way are highly porous ($\phi < 10^{-3}$) and mimic those that form during the early stages of dust growth at low collision speeds {\citep[e.g.,][]{Ormel+07,Okuzumi+09,Okuzumi+12,Zsom+10}}. 
Next, we compress each BCCA aggregate sample until its filling factor reaches a designated value (0.4, 0.45, or 0.5) by moving periodic boundaries, following the procedure developed by \citet{Tatsuuma+23}.
We then connect some identical compressed BCCA as needed and cut out a sphere of radius $R_{\rm agg}$.
Finally, we allow the monomers within each of the four spherical aggregate samples to move under an artificial damping force, in addition to the monomer interaction forces described in Section~\ref{sec:interaction_model}, until the number of contact points within each sample relaxes into a constant value.
This artificial damping force is given by Equation (5) of \citet{Tatsuuma+19}, with the dimensionless damping coefficient $k_\mathrm{n}$ set to 0.01.
The changes in aggregate radius and mass through this final relaxation process are negligible.

\begin{table}
    \centering
    \caption{Initial Volume Filling Factors and Radii of Spherical Aggregates Used in the Simulations}
    \begin{tabular}{cc} \hline \hline
         Volume filling factor $\phi$ & Aggregate radius $R_{\mathrm{agg}}/r_1$ \\ \hline
        0.4 & $30,~40,~\ldots, 120$ \\
        0.45 & $30,~40,~\ldots, 70$ \\
        0.5 & $20,~30,~\ldots, 70$ \\ \hline
    \end{tabular}
    \label{tab:agg_params}
\end{table}

Table \ref{tab:agg_params} lists the volume filling factors and aggregate radii of the aggregates created in the above procedure. 
Each aggregate consists of $N_{\mathrm{agg}}\simeq (R_{\mathrm{agg}}/r_1)^3\phi$ monomers. 
In our simulations, $N_{\mathrm{agg}}$ is between about 4,000 and 700,000.
The aggregate mass can be approximately evaluated as
\begin{equation}
m_{\mathrm{agg}}=m_1N_{\mathrm{agg}}\simeq m_1\left(\frac{R_{\mathrm{agg}}}{r_1}\right)^3\phi.
\end{equation} 

\begin{figure}[t]
    \centering
    \includegraphics[width=0.45\textwidth]{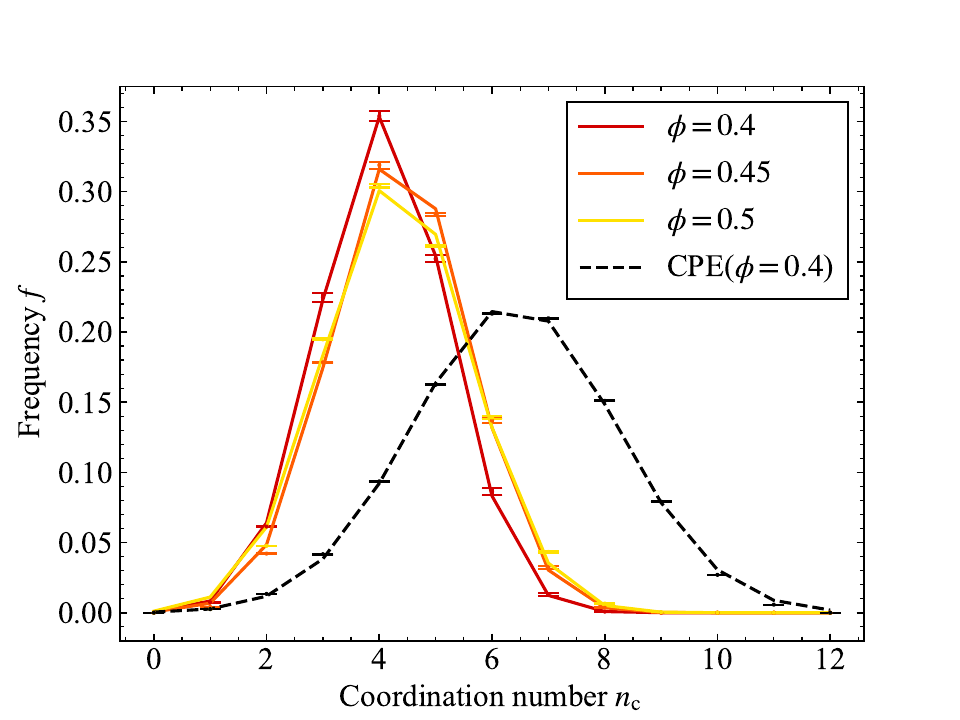}
    \caption{Frequency distribution of the coordination numbers of monomers in compressed BCCA spheres (solid lines) and CPE spheres (dashed line)
    with $R_{\mathrm{agg}}/r_1=70$. 
    Lines are Gaussian fits. Vertical error bars represent the standard derivations. 
    }
    \label{fig:nc}
\end{figure}

\begin{figure*}[t]
    \centering
    \includegraphics[width=\textwidth]{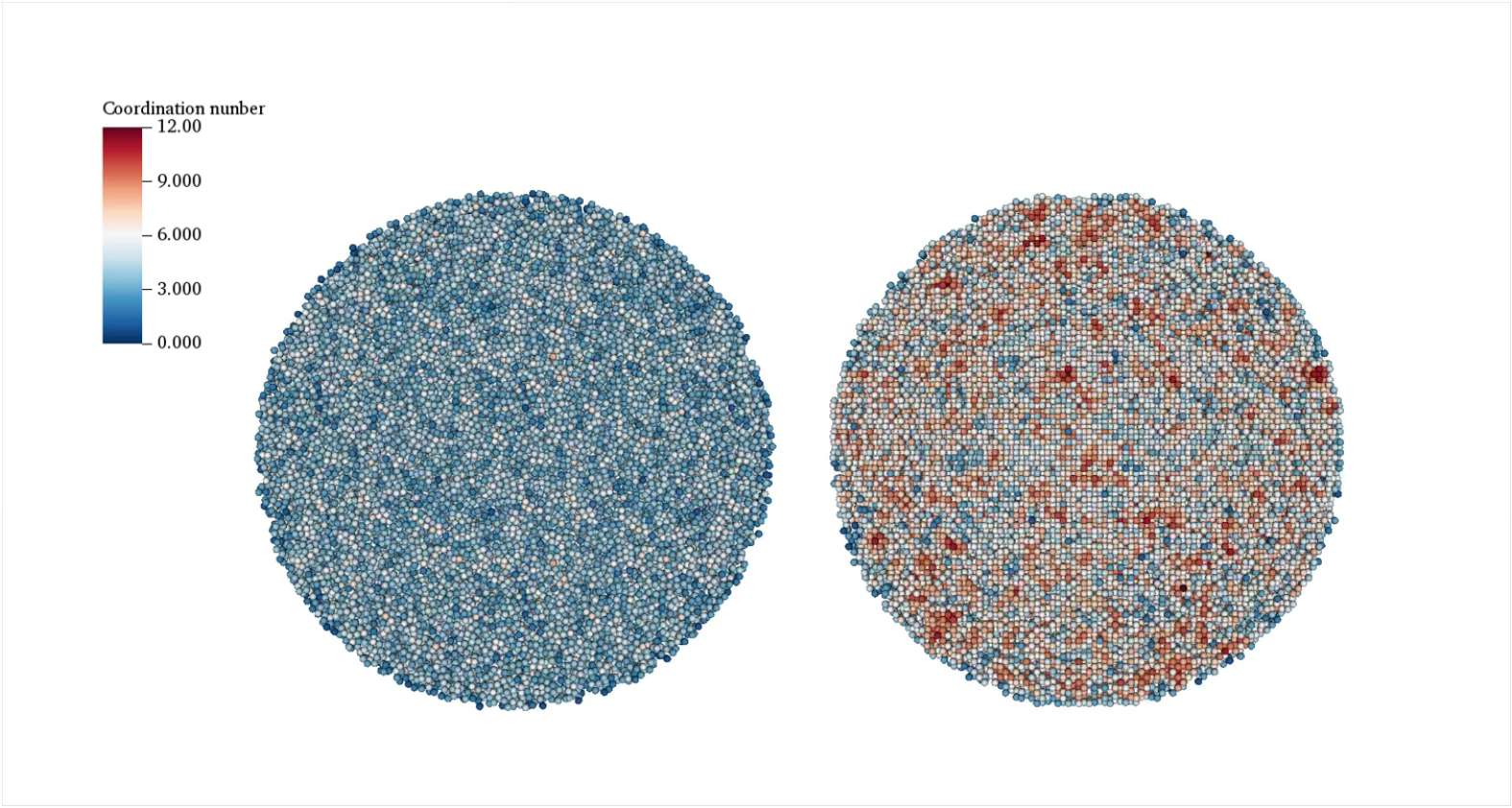}
    \caption{Cross section of a compressed BCCA sphere (left) and a CPE sphere (right) with $\phi=0.4$ and $R_{\mathrm{agg}}/r_1=70$. 
    The color represents the coordination number of each particle, with blue indicating a lower value and red indicating a higher value. }
    \label{fig:cross-section}
\end{figure*} 

Figure \ref{fig:nc} shows the frequency distribution of the coordination numbers of monomers in different types of aggregate (compressed BCCAs with $\phi=0.4$, 0.45, and 0.5 and CPE spheres with $\phi=0.4$) of an equal aggregate size of $R_{\mathrm{agg}}/r_1=70$, obtained using four aggregate samples for each type.  
The average coordination numbers of the compressed BCCA spheres with $\phi=0.4,~0.45$, and 0.5 and CPE spheres are $4.10,~4.37,~4.32$ and $6.40$, respectively. 
\citet{Arakawa+19} showed that the average coordination number of compressed BCCA is related to $\phi$ as
\begin{equation} \label{eq:phi-nc}
    \bar{n}_{\mathrm{c}}(\phi)=2+9.38\phi^{1.62}.
\end{equation}
This equation predicts that the average coordination numbers of the compressed BCCA with $\phi=0.4$, $0.45$, and $0.5$ are $4.18$, $4.64$, and $5.13$, respectively. 
The values for our aggregates are smaller than the prediction from Equation (\ref{eq:phi-nc}) because our aggregates have separated some contact points during damping. 
Figure \ref{fig:cross-section} shows the cross section of a compressed BCCA sphere and a CPE sphere with $\phi=0.4$. 
Almost all monomers in the compressed BCCA aggregate have coordination numbers less than $6$. 
In contrast, the CPE sphere has locally concentrated regions of coordination numbers$\gtrsim 9$, which occupy $\approx 10\%$ of the entire aggregate volume.

\subsection{Collision Simulations} \label{sec:col_sim}
The impact velocity of two dust aggregates is set to $v_{\mathrm{imp}}=2\times10^{(0.1i)} ~\mathrm{m\,s^{-1}}$, where $i=0,~1,~\ldots,~10$. 
For each parameter set $(\phi, R_{\mathrm{agg}}, v_{\mathrm{imp}})$, we simulate a head-on collision between two spherical aggregates with varying orientations.
The end time $t_{\rm end}$ of each calculation is determined according to 
\begin{equation}
    t_{\mathrm{end}}=6.4\times10^3 \left(\frac{R_{\mathrm{agg}}}{r_1}\right)\left(\frac{v_{\mathrm{imp}}}{1~\mathrm{m~s^{-1}}}\right)^{-1}t_{\mathrm{c}} . 
\end{equation}

\section{Collision Outcomes and Bouncing Condition} \label{sec:result}

In this section, we describe the collision outcomes with the fiducial parameters to investigate the impact velocity dependency (Section \ref{sec:fiducial}). 
We then discuss the relationship between the aggregate mass and impact velocity where bouncing occurs and its dependence on the filling factor in Section \ref{sec:result_params}.

\subsection{Fiducial Runs} \label{sec:fiducial}

\begin{figure*}[t]
    \centering
    \includegraphics[width=\textwidth]{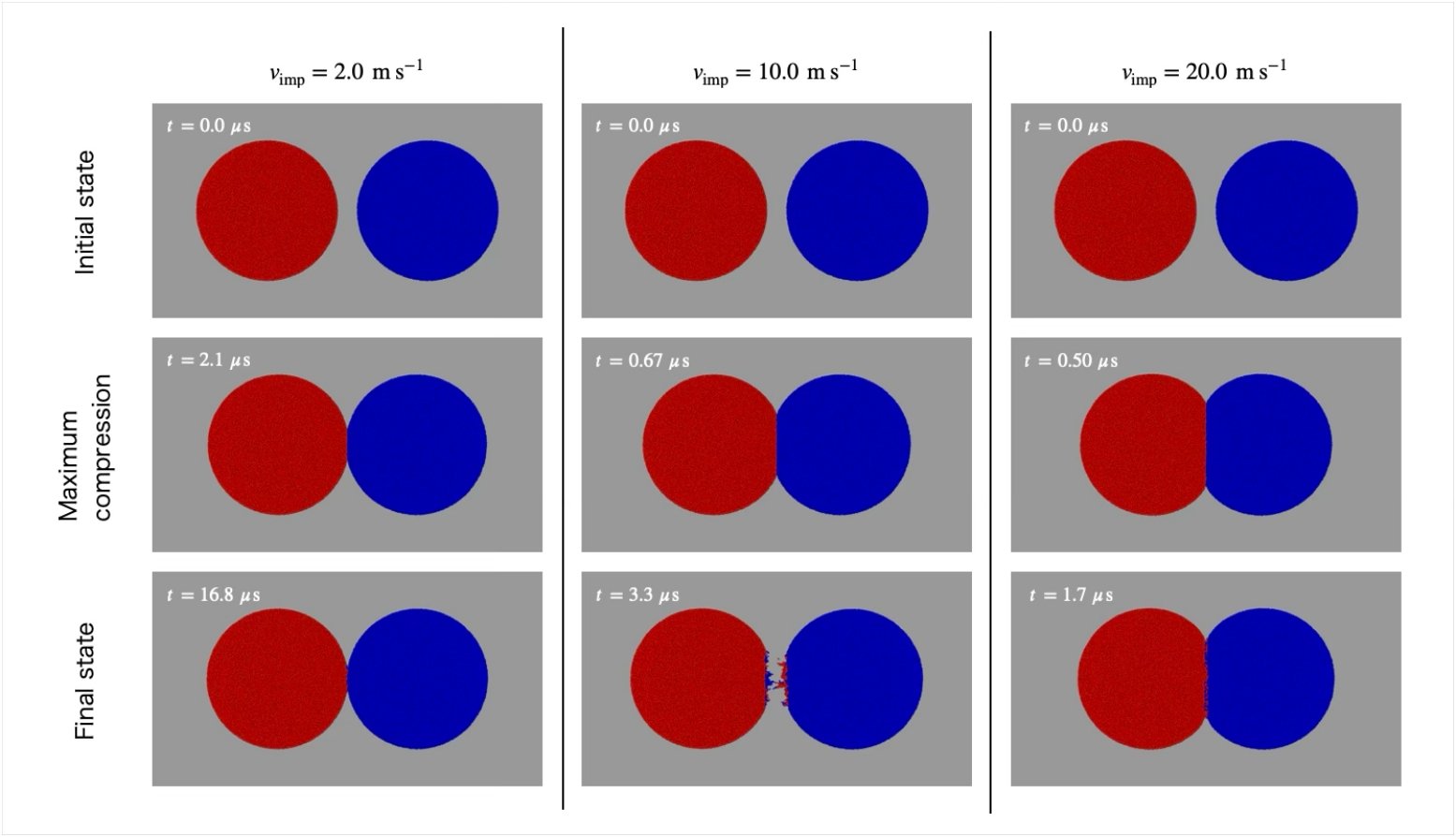}
    \caption{Snapshots of three runs with $v_{\mathrm{imp}}=2,~10,$ and $20~\mathrm{m\,s^{-1}}$ for $\phi=0.4$ and  $R_{\mathrm{agg}}/r_1=110$.
    The top, middle, and bottom rows show the initial, maximally compressed , and final states, respectively. 
    The initial aggregates are the same for all the illustrated runs.}
    \label{fig:snapshot}
\end{figure*}

To begin with, we present results from simulation runs for the fiducial parameter set of $\phi=0.4$ and $R_{\mathrm{agg}}/r_1=110$ to illustrate how the collision outcome depends on impact velocity $v_{\mathrm{imp}}$. 
Figure \ref{fig:snapshot} shows snapshots from runs with $v_{\mathrm{imp}}=2$, 10, and 20$~\mathrm{m~s^{-1}}$. 
The initial aggregates are the same in these runs. 
The aggregates bounce at the intermediate velocity of $v_{\mathrm{imp}}=10~\mathrm{m~s^{-1}}$. 
They stick at $v_{\mathrm{imp}}=2$ and 20$~\mathrm{m~s^{-1}}$.
This impact velocity dependence is discussed below.

\begin{figure}[t]
    \centering
    \includegraphics[width=0.45\textwidth]{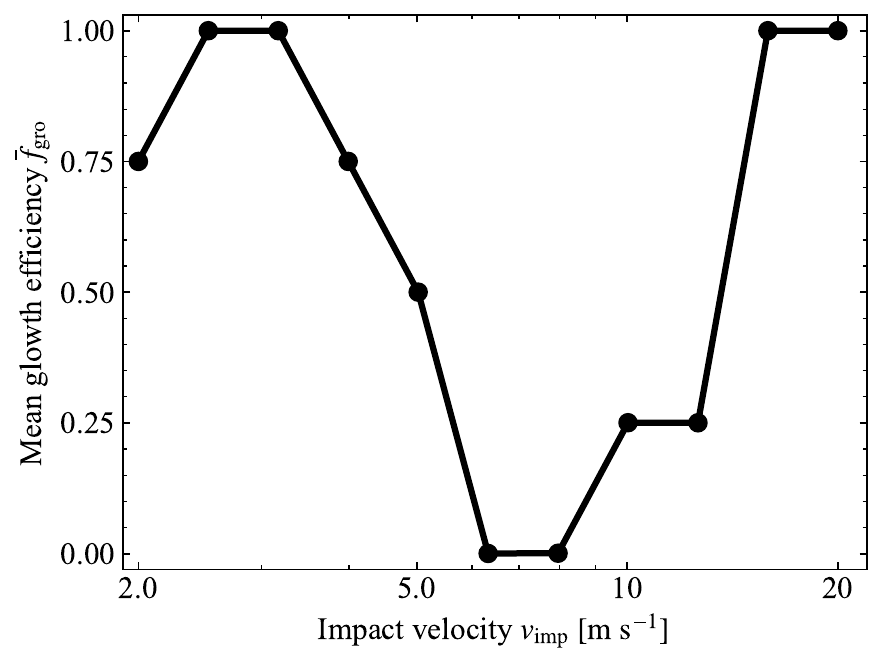}
    \caption{Mean collisional growth efficiency as a function of impact velocity for runs with $\phi=0.4$ and $R_{\mathrm{agg}}/r_1=110$.}
    \label{fig:fiducial_result}
\end{figure}

In order to quantify the collision outcome, we use the collisional growth efficiency $f_\mathrm{gro}$ for an individual collision, defined by \citep{Wada+13,Arakawa+23}
\begin{equation}
    f_{\mathrm{gro}}=\frac{N_{\mathrm{large}}-N_{\mathrm{agg}}}{N_{\mathrm{agg}}},
\end{equation}
where 
$N_{\mathrm{large}}$ is the number of monomers involved in the largest aggregate that forms after collision. 
The growth efficiency defined above is unity for perfect sticking and vanishes for perfect bouncing of equal-sized aggregates without mass transfer.
All of our runs result in either near-perfect sticking ($f_{\mathrm{gro}}\approx 1$) or near-perfect bouncing ($f_{\mathrm{gro}}\approx 0$), depending on internal BCCA structure of the colliding aggregates as well as the parameter set $(\phi,R_{\mathrm{agg}},v_{\mathrm{imp}})$.

We also define the mean growth efficiency $\bar{f}_{\mathrm{gro}}$ as the average of $f_{\mathrm{gro}}$ over four runs with the same $(\phi,R_{\mathrm{agg}},v_{\mathrm{imp}})$ but with different initial aggregate samples.
We define the threshold between sticking and bouncing collisions as $\bar{f}_{\mathrm{gro}}=0.5$.

To discuss the dependence of collision outcome on impact velocity, we plot the mean growth efficiency as a function of impact velocity in Figure \ref{fig:fiducial_result}. 
The result indicates that intermediate impact velocity $(5~\mathrm{m\,s^{-1}}\lesssim v_{\mathrm{imp}}\lesssim15~\mathrm{m\,s^{-1}})$ is required for bouncing for the aggregates with fiducial parameters $\phi=0.4$ and $R_{\mathrm{agg}}/r_1=110$.
Low velocity collisions and high velocity collisions cause sticking for different reasons: at the lower velocity, sticking is simply due to the low impact energy, while at the higher velocity, sticking results from significant deformation of the aggregates upon impact, leading to adhesion through a large contact area. 
This impact velocity dependence was not found in previous simulations \citep{Arakawa+23}.

\subsection{Parameter Dependence} \label{sec:result_params}

\begin{figure*}[t]
    \centering
    \includegraphics[width=\textwidth]{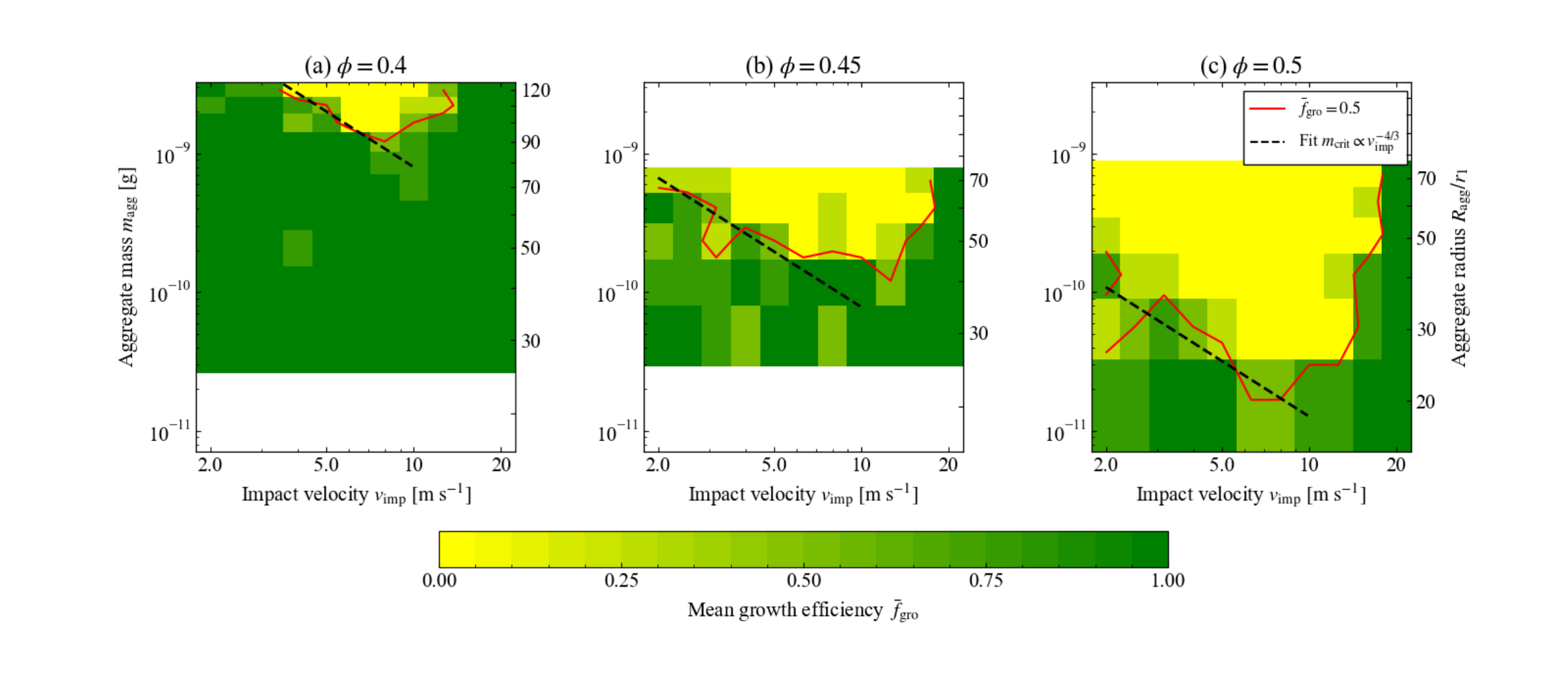}
    \caption{Mean growth efficiency as a function of impact velocity and aggregate mass (or radius) for three values of $\phi$, plotted across the parameter regions explored in this study. 
    The values $\bar{f}_{\rm gro} = 1$ and 0 correspond to perfect sticking and bouncing for all four runs, respectively. 
    Solid lines represent the sticking--bouncing threshold, where $\bar{f}_{\mathrm{gro}}=0.5$.
    Dashed lines represent power-law fits (Equation (\ref{eq:fitting_phimv1})).}
    \label{fig:result_mv}
\end{figure*}

We now discuss how the outcome of aggregate collisions depends on impact velocity and aggregate size. 
Figure \ref{fig:result_mv} displays the mean collisional growth efficiency $\bar{f}_{\rm gro}$ as a function of $v_{\mathrm{imp}}$ and $R_{\mathrm{agg}}$ for all volume filling factors explored in this study. 
In the results of $\phi=0.4$ (Figure~\ref{fig:result_mv} (a)), small aggregates $R_{\mathrm{agg}}/r\lesssim80$ stick in the entire velocity range explored in this study ($v_{\rm imp} =$ 2--20 $\rm m~s^{-1}$).
As $R_{\rm agg}$ exceeds a certain value, bouncing starts to occur at intermediate impact velocities. 
A similar trend can be observed in the results for $\phi=0.45$ and 0.5 (Figure~\ref{fig:result_mv} (b) and (c), respectively). 
The critical aggregate size above which bouncing can occur depends strongly on the volume filling factor. 
This dependence is discussed below.

At low impact velocities ($v_{\rm imp} \lesssim 10~\rm m~s^{-1}$), we observe a trend that larger aggregates bounce at lower $v_{\rm imp}$.
This trend is qualitatively consistent with the results of laboratory experiments \citep[e.g.,][]{Guttler+10,Weidling+12,Kothe+13}. 
\citet{Kothe+13} showed that the threshold mass $m_{\rm crit}$ above which bouncing occurs follows a power-law relation $m_{\rm crit} \propto v_{\mathrm{imp}}^{-4/3}$.

To allow for a direct comparison with the experimental results, we replot $\bar{f}_{\rm gro}$ as a function of $v_{\mathrm{imp}}$ and $m_{\mathrm{agg}}$ in Figure~\ref{fig:result_mv}.
The solid lines show the threshold mass $m_{\rm crit}$, where $\bar{f}_{\rm gro} = 0.5$, obtained by interpolating our data. 
For the threshold mass curves at $v_{\mathrm{imp}}<10~\mathrm{m\,s^{-1}}$, we perform a least-square fitting of a power-law function of the form
\begin{equation}
    \log_{10}{m_{\mathrm{bounce}}}=b-\frac{4}{3}\log_{10}{\left(\frac{v_{\mathrm{imp}}}{1~\mathrm{m\,s^{-1}}}\right)},
\end{equation}
where $b$ is a fitting parameter. 
This fitting function can also be written as
\begin{equation} \label{eq:fitting_phimv1}
    m_{\mathrm{bounce}}=m_{v1}\left(\frac{v_{\mathrm{imp}}}{1~\mathrm{m\,s^{-1}}}\right)^{-4/3},
\end{equation}
where $m_{v1}$ stands for the threshold mass at $v_{\mathrm{imp}}=1~\mathrm{m\,s^{-1}}$.
Equation~\eqref{eq:fitting_phimv1} can also be represented as 
\begin{equation}
v_{\rm bounce} = 1\pfrac{m_{\rm agg}}{m_{v1}}^{-3/4}~\rm m~s^{-1},
\label{eq:v_bounce}
\end{equation}
where $v_{\rm bounce}$ stands for the threshold impact velocity for bouncing.
Table~\ref{tab:fitting_result} shows the best-fit $m_{v1}$ values for three values of $\phi$. 
The best-fit functions are shown by the dashed lines in Figure \ref{fig:result_mv}. 
We find that the best-fit lines reproduce the overall trend of the threshold mass curves at $v_{\mathrm{imp}} \lesssim 10~\mathrm{m\,s^{-1}}$ directly obtained from our simulations.

\begin{table*}[t]
    \centering
    \caption{Best-fit Values of $m_{v1}$.}
    \begin{tabular}{ccc} \hline \hline
       Volume filling factor $\phi$ & Fitting parameter $m_{v1}~[\mathrm{g}]$ & $\log_{10}{m_{v1}}$ 1-$\sigma$ scatter $[\mathrm{g}]$\\ \hline
       $0.4$ & $1.67\times10^{-8}$ & $7.12\times10^{-10}$ \\
       $0.45$ & $1.68\times10^{-9}$ & $1.64\times10^{-10}$\\
       $0.5$ & $2.74\times10^{-10}$ & $3.79\times10^{-11}$\\ \hline
    \end{tabular}
    \tablecomments{See Equation~(\ref{eq:fitting_phimv1}) for the fitting function.}
    \label{tab:fitting_result}
\end{table*}

\begin{figure}[t]
    \centering
    \includegraphics[width=0.45\textwidth]{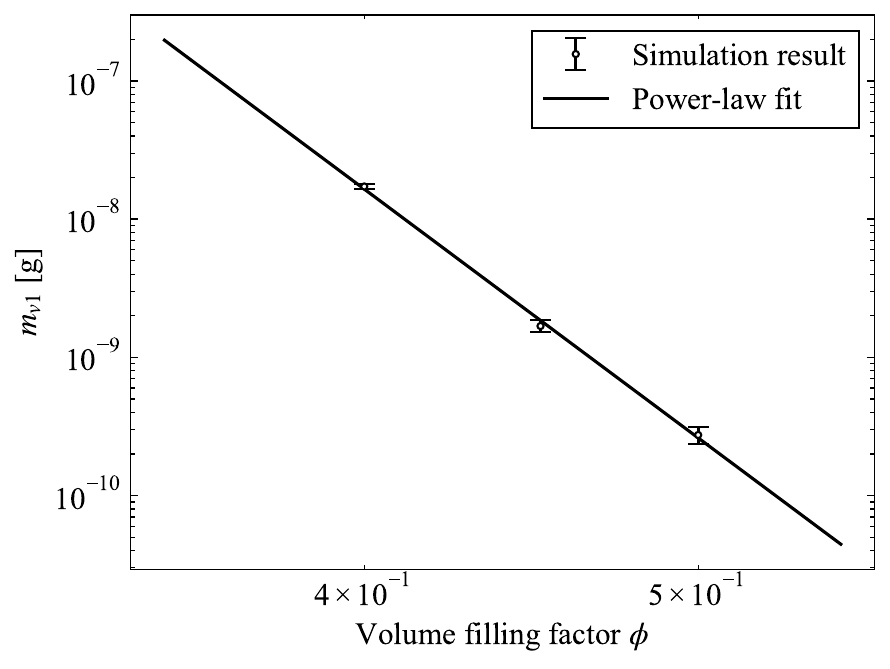}
    \caption{Best-fit values of $m_{v1}$ in the empirical power-law relation between $m_{\rm crit}$ and $v_{\rm imp}$ (Equation~\eqref{eq:fitting_phimv1}) for $\phi = 0.4$, 0.45, and 0.5 (circle symbols). 
    The error bars represent the 1-$\sigma$ scatters of $m_{\rm crit}$ from the simulations (solid lines in Figure~\ref{fig:result_mv}) around the best-fit curve (dashed lines in Figure~\ref{fig:result_mv}). 
    The line indicates a power-law fit $m_{v1} \propto \phi^{-18.6}$. See also Table \ref{tab:fitting_result} for the values of $m_{v1}$ and 1-$\sigma$ scatters. }
    \label{fig:result_phimv1}
\end{figure}

The threshold aggregate mass for bouncing is highly sensitive to the volume filling factor. 
Figure \ref{fig:result_mv} indicates that the threshold mass decreases by almost an order of magnitude {when $\phi$ is increased by 0.05}. 
Figure \ref{fig:result_phimv1} shows $m_{v1}$ as a function $\phi$. 
Power-law fitting shows that $m_{v1}$ scales with $\phi$ as steeply as $\phi^{-18.6}$. 
We discuss one possible interpretation of this strong dependence in Section \ref{sec:Pcomp}.

\section{Aggregate Collision Energetics} \label{sec:energy}

Since the bouncing of colliding aggregates is caused by elastic repulsion, knowing the fractions of initial impact energy converted into elastic potential energy and how energy dissipates 
is essential for fully understanding the origin of the bouncing/sticking threshold. 
In this section, we present an in-depth analysis on the energy conversion processes in our collision simulations.

\subsection{Overview and Definitions} \label{sec:recipe}

\begin{figure*}[t]
    \centering
    \includegraphics[width=\textwidth]{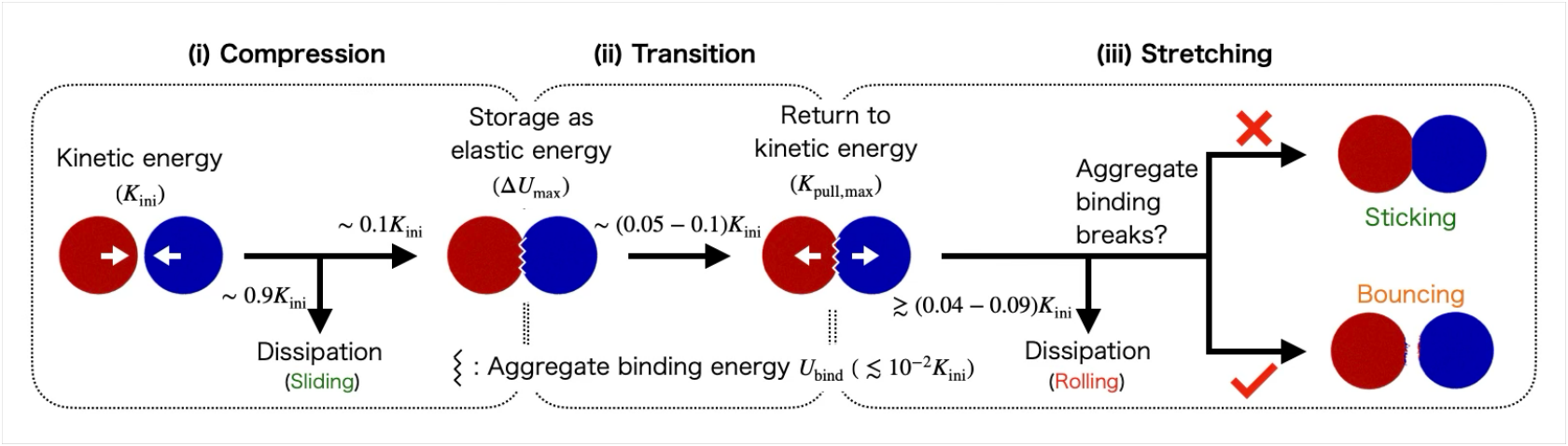}
    \caption{Schematic showing three phases of non-fragmenting aggregate collisions. 
    The arrows indicate energy conversion flows and collision outcomes, with $K_{\mathrm{ini}},~\Delta U_{\mathrm{max}},~K_{\mathrm{pull,max}},~U_\mathrm{bind}$ denoting the initial impact energy, aggregate elastic energy at the maximum compression, maximum kinetic energy after maximum compression, aggregate binding energy, respectively.}
    \label{fig:col_model}
\end{figure*}

We analyze the energy flows during an aggregate collision by dividing the event into three phases (Figure \ref{fig:col_model}). 
In the initial compression phase, colliding aggregates compress each other, dissipating some of the impact energy through internal plastic deformation while storing the rest as elastic energy. 
This phase is followed by the transition phase, where the compressed aggregates repel each other, converting the stored elastic energy back into kinetic energy. 
In the final stretching phase, part of this kinetic energy is dissipated through plastic deformation, and the repelling aggregates either bounce off or remain in contact, depending on the amount of remaining kinetic energy. 
This section aims to quantify these energy flows based on our simulations.

Below, we define quantities needed to formulate the collision energetics. 
We consider a head-on collision of two equal-sized spherical aggregates 1 and 2 along the $x$-axis. We define the compressive distance of the aggregates as
\begin{equation}
    \delta_{\mathrm{agg}}\equiv 2R_{\mathrm{agg}}-|x_{\mathrm{agg,1}}-x_{\mathrm{agg,2}}|,
\end{equation}
where $x_{\mathrm{agg,1}}$ and $x_{\mathrm{agg,2}}$ are the $x$-components of the centers of mass of aggregates 1 and 2
\footnote{More precisely, the centers of mass of all monomers that originally belonged to aggregates 1 and 2. However, the fragments from the colliding aggregates carry negligible mass and therefore have little effect on $x_{\mathrm{agg,1}}$ and $x_{\mathrm{agg,2}}$.}, 
respectively.
We then define the aggregate kinetic energy as the kinetic energy of the relative motion between the aggregates' centers of mass,
\begin{align}
    K_{\mathrm{agg}}&=\frac{1}{2}\frac{m_{\mathrm{agg,1}}m_{\mathrm{agg,2}}}{m_{\mathrm{agg,1}}+m_{\mathrm{agg,2}}}\left(\frac{d\delta_{\mathrm{agg}}}{dt}\right)^2 \notag \\
    &=\frac{m_{\mathrm{agg}}}{4}\left(\frac{d\delta_{\mathrm{agg}}}{dt}\right)^2,
\end{align}
where $m_{\mathrm{agg1,2}}(=m_{\mathrm{agg}})$ represents the mass of each aggregate{, and $t$ represents time}. 
In particular, we express the initial impact energy as $K_\mathrm{ini}(=m_{\mathrm{agg}}v^2_{\mathrm{imp}}/4)$. 
We note that $K_{\rm agg}$ does not include kinetic energy associated with the internal deformation (e.g., vibrations) of the aggregates.

The aggregate elastic energy $U$ is defined as the total elastic energy of the monomers composing the colliding aggregates, and we consider its time variation $\Delta U(t) = U(t)-U(0)${, where $U(0)$ is the aggregate elastic energy at the initial state}.
While the elastic energy of a monomer can be divided into the elastic energy associated with normal motion and tangential motion, we consider only the normal elastic energy because the tangential elastic energy is negligibly small. 
The normal elastic energy $U_{\mathrm{n}}$ is defined as Equation (6) in \citet{Wada+07}, but we set $U_{\mathrm{n}}=0$ when $\delta=0$ so that the aggregate elastic energy does not change when the coordination number changes. 

We also define the binding energy of the aggregates after contact as 
\begin{equation} 
    U_{\mathrm{bind}}=n_\mathrm{c,agg}E_{\mathrm{break}},
\end{equation}
where $n_\mathrm{c,agg}$ is the number of contact points between the monomers that originally belonged to different aggregates. 

The cumulative dissipated energy is denoted by $E_{\mathrm{dis}}$. In particular, the subscript ``pull'' refers to the cumulative dissipated energy after maximum compression.

\subsection{Analysis of Fiducial Runs} \label{sec:energy_fiducial}

\begin{figure*}[t]
    \centering
    \includegraphics[width=\textwidth]{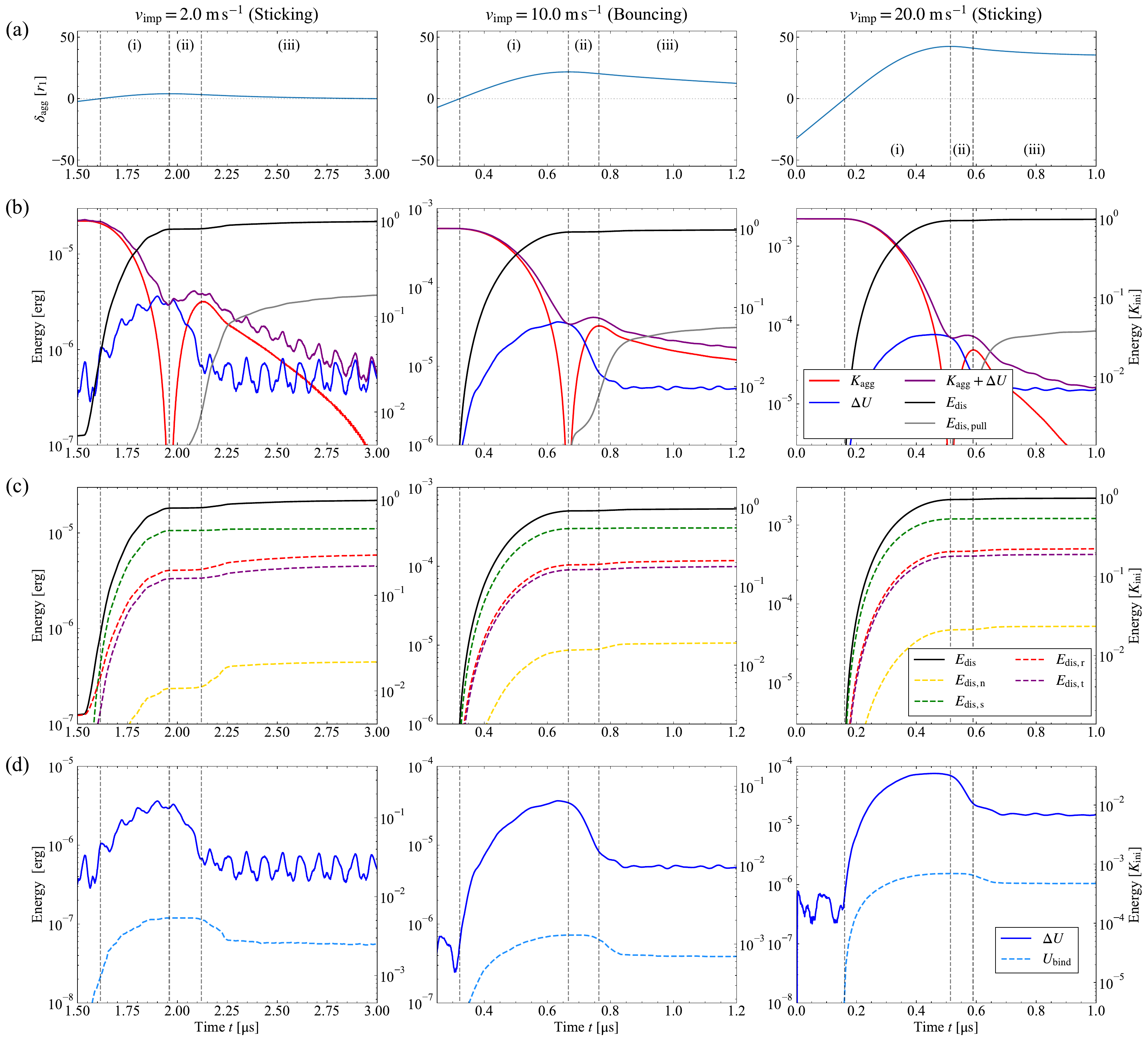}
    \caption{Aggregate compressive length $\delta_{\rm agg}$ (rows (a))  and  energies (rows (b)--(d)) as a function of time {$t$} from fiducial runs with $v_{\mathrm{imp}}=2,~10,$ and $20~\mathrm{m\,s^{-1}}$ (left, center, and right columns, respectively). 
    In each panel, the leftmost vertical line indicates the time of contact formation, and the other two vertical lines separate the compression, transition, and stretching phases.
    In panels (b), the red, blue, {purple,} black, and gray lines represent the aggregate kinetic energy $K_{\mathrm{agg}}$, aggregate elastic energy $\Delta U$, {their sum $K_{\mathrm{agg}}+\Delta U$}, cumulative dissipation energy $E_{\mathrm{dis}}$, and cumulative dissipation energy after maximum compression $E_{\mathrm{dis,pull}}$, respectively.
    In panels (c), the yellow, green, red, and purple lines display the cumulative dissipation energies caused by normal $E_{\mathrm{dis,n}}$, sliding $E_{\mathrm{dis,s}}$, rolling $E_{\mathrm{dis,r}}$, and twisting motions $E_{\mathrm{dis,t}}$ displacements, respectively, together with the total dissipated energy $E_{\mathrm{dis}}$ indicated by the black line.
    In panels (d), the solid and dashed lines show the aggregate elastic energy $\Delta U$ and aggregate binding energy $U_{\mathrm{bind}}$.}
    \label{fig:Energy_all}
\end{figure*}

\begin{figure*}[t]
    \centering
    \includegraphics[width=\textwidth]{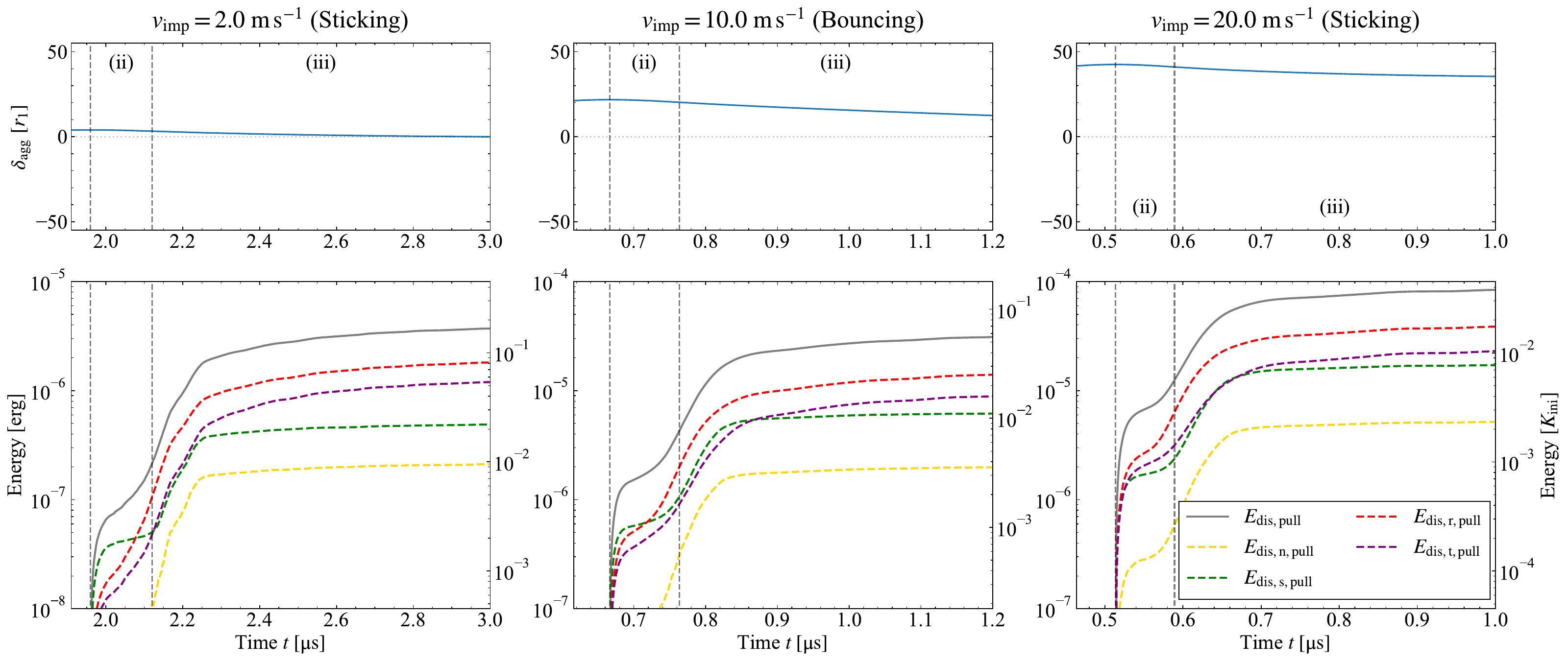}
    \caption{Same as panels (a) and (c) in Figure \ref{fig:Energy_all} (upper and lower panels, respectively), but the bottom panels show the cumulative dissipation energy after maximum compression. 
    The total cumulative dissipation energy after maximum compression is shown by the gray line.}
    \label{fig:edis_pull}
\end{figure*}

To begin with, we analyze the collision energetics in the three runs with the fiducial parameters $\phi=0.4$, $R_{\mathrm{agg}}/r_1=110$, $v_{\mathrm{imp}}=2$, 10, and 20 $\mathrm{m\,s^{-1}}$ shown in Figure~\ref{fig:snapshot}.
Figure \ref{fig:Energy_all} shows the time variation of aggregate compressive length and energies 
{\footnote{Note that Figures \ref{fig:Energy_all} and \ref{fig:edis_pull} do not include data up to the end time of calculation $t_{\rm end}\approx25$, 5, and $2.5~\mathrm{\mu s}$ for $v_{\rm imp}=2$, 10, and 20 $\rm m\,s^{-1}$ cases, respectively.}}.

\begin{figure*}[t]
    \centering
    \includegraphics[width=\textwidth]{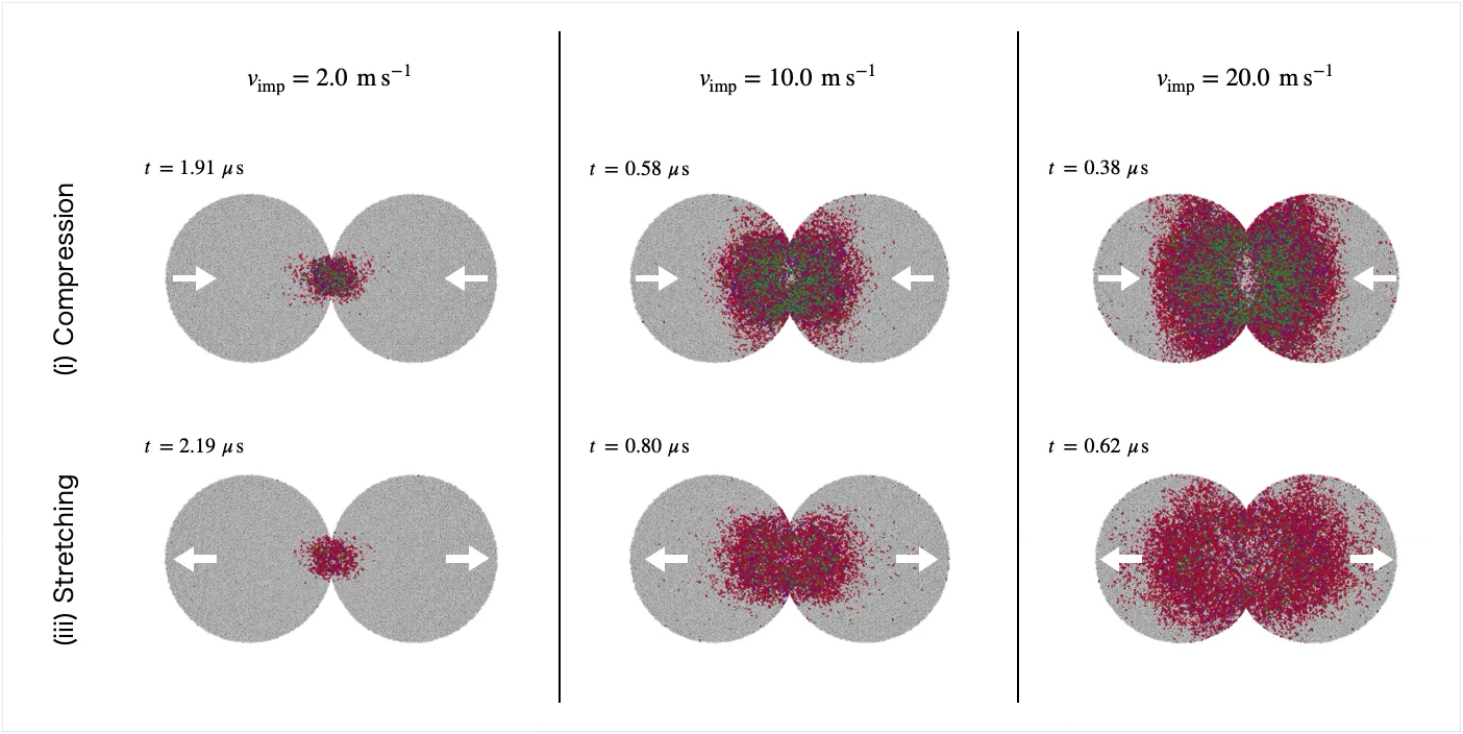}
    \caption{Cross section of colliding aggregates from fiducial runs, with $v_{\mathrm{imp}}=2,~10,$ and $20~\mathrm{m\,s^{-1}}$, showing the spatial distribution of monomers dissipating energy in the compression and stretching phases. 
    The colors indicate the dominant energy dissipation processes: green for sliding, red for rolling, and purple for twisting.}
    \label{fig:edis_cross}
\end{figure*}

In the initial compression phase (region (i) in Figure~\ref{fig:Energy_all}), only approximately 10\% of the initial impact kinetic energy $K_{\rm agg}$ is converted into elastic energy $\Delta U$, while the remaining 90\% is dissipated during the compression process (Figure~\ref{fig:Energy_all}(b)). 
The main energy dissipation process is sliding, followed by rolling, twisting, and normal motion (Figure~\ref{fig:Energy_all}(c)). 
This trend is consistent with the energy dissipation during static compression of aggregates \citep{Tatsuuma+23}. 
These energy dissipations occur in a spherical region surrounding the contact surface (Figure~\ref{fig:edis_cross}).
The aggregate binding energy $U_{\rm bind}$ increased during the compression phase is an order of magnitude smaller than $\Delta U$ and two orders of magnitude smaller than $K_{\rm ini}$ (Figure~\ref{fig:Energy_all}(d)).

In the transition phase (region~(ii) in Figure~\ref{fig:Energy_all}), nearly all of the stored elastic energy is converted back to kinetic energy, with negligible energy dissipation. 
The parameter dependence of this conversion efficiency is discussed in Section~\ref{sec:energy_params}.

During the stretching phase (region~(iii) in Figure~\ref{fig:Energy_all}), about 70\% of the sum of the converted kinetic energy and elastic energy is dissipated in the bouncing case and more than 95\% in the sticking case (Figure~\ref{fig:Energy_all}(b)). 
\footnote{We note that more energy than the sum of converted aggregate kinetic energy and aggregate elastic energy can be dissipated because some energy can be dissipated by vibrations of monomers that do not contribute to the aggregate motion.}
{In the bouncing case with $v_{\rm imp} = 10~\rm m\,s^{-1}$ (middle panel of Figure~\ref{fig:Energy_all}), for example, the sum $K_{\rm agg}+\Delta U$ is $\approx 0.08K_{\rm ini}$ at the beginning of the stretching phase ($t\approx 0.76~\rm \mu s$), whereas the cumulative dissipation energy after maximum compression, $E_{\rm dis,\,pull}$, reaches $\approx0.055K_{\rm ini}$ at the end of the simulation ($t\approx 5.0~\rm \mu s$), so that $\approx 0.055/0.08 \approx$ 70\% of the available energy has been dissipated.}
This difference can be attributed to the fact that in the sticking case the aggregate continues to dissipate energy while expanding and contracting until the end time, while in the bouncing case it dissipates almost no energy after breaking the contact surface. 
Therefore, this difference in values depends on the collision outcome, but does not determine the collision outcome.
The main dissipation process in this phase is rolling, followed by twisting, sliding, and normal motion (Figure~\ref{fig:edis_pull}).
This trend differs from that in the compression phase but is consistent with the energy dissipation during static stretching of aggregates \citep{Tatsuuma+19}.

\subsection{Parameter Dependence of Energy Conversion} \label{sec:energy_params}

\begin{figure*}[t]
    \centering
    \includegraphics[width=\textwidth]{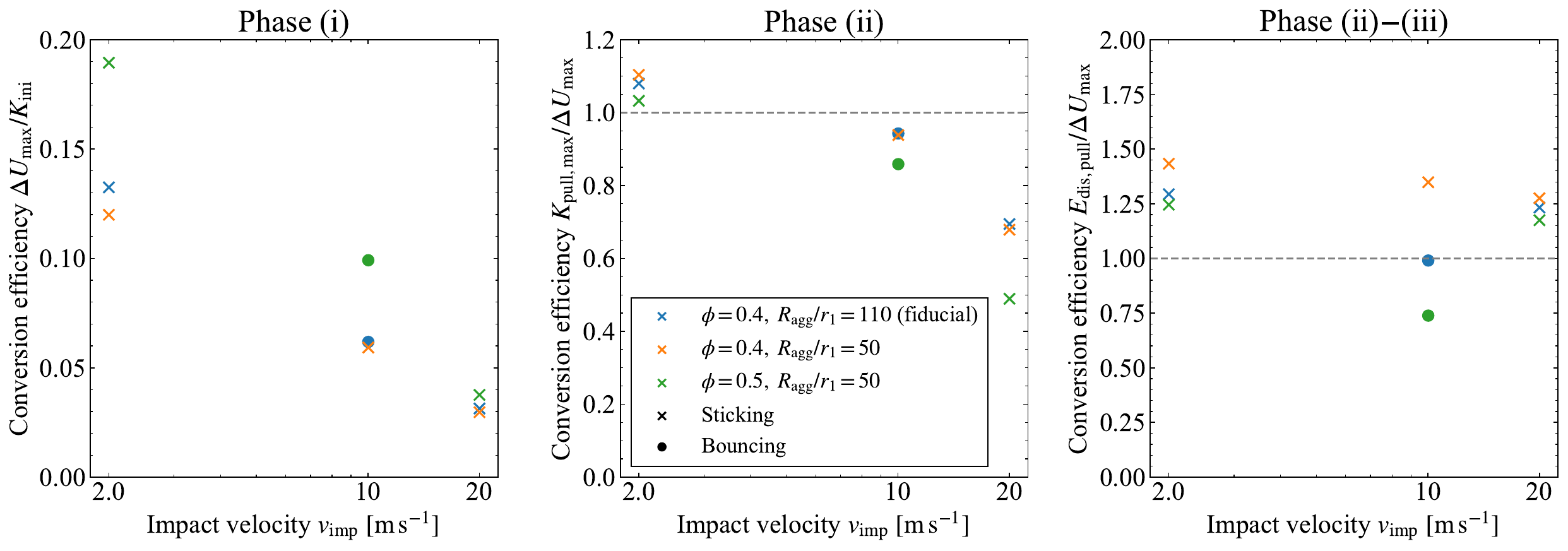}
    \caption{Efficiencies of conversion between the aggregate kinetic energy $K_{\mathrm{agg}}$ and aggregate elastic energy $\Delta U$ during the compression and transition phases (left and middle panels, respectively), and the energy dissipation rate of the stored aggregate elastic energy (right panel) against impact velocity, for aggregates of different $(\phi,R_{\mathrm{agg}})$. Sticking and bouncing cases are denoted by cross and circular symbols, respectively. } 
    \label{fig:energy_eff}
\end{figure*}

We {now} turn our attention to the parameter dependence of energy conversion.
The left panel of Figure \ref{fig:energy_eff} shows the dependencies of the energy conversion efficiency during the compression phase, defined as the fraction of kinetic energy $K_{\rm agg}$ going to elastic energy $\Delta U$, on the parameters $(\phi, R_{\rm agg}, v_{\rm imp})$. 
The conversion efficiency decreases with increasing impact velocity. 
However, its dependence is weak: even if the impact velocity is increased by a factor of $5$ from $v_{\mathrm{imp}}=2~\mathrm{m\,s^{-1}}$ to $10~\mathrm{m\,s^{-1}}$, the efficiency is reduced only by a factor of 2.
The velocity dependence of the energy conversion efficiency is stronger for aggregates of higher $\phi$. This is because the internal restructuring of more compact aggregates is more difficult at low impact energies.

Energy conversion during the transition phase, defined as the fraction of elastic energy at maximum compression going to maximum kinetic energy after maximum compression, is efficient.
The middle panel of Figure \ref{fig:energy_eff} shows that the energy conversion efficiency during this phase {$K_{\rm pull,\,max}/\Delta U_{\rm max}$} is close to unity at $v_{\mathrm{imp}} \lesssim 10~\mathrm{m\,s^{-1}}$
\footnote{We note that the conversion efficiency defined here can slightly exceed unity because the elastic energy is not always maximized at the moment of maximum compression.}. 
As the impact velocity increases to $20~\rm m~s^{-1}$, the conversion efficiency decreases to 0.5--0.7, indicating that energy dissipation during the transition phase becomes significant at such high velocities. This significant energy dissipation may explain aggregate sticking observed at $v_{\rm imp} = 20~\rm m~s^{-1}$. 

The rate of energy dissipation after maximum compression {$E_{\rm dis,\,pull}/\Delta U_{\rm max}$} depends on the collision outcomes.
The right panel of Figure \ref{fig:energy_eff} shows that the dissipation ratio is less than unity for bouncing cases, while it exceeds unity for sticking cases \footnote{In the right panel, aggregates dissipate energy as they oscillate in sticking cases. To exclude extra dissipation, we use $E_{\mathrm{dis,pull}}$ here as the cumulative dissipated energy from the maximum compression until the aggregate returns to the compressive direction (i.e., $d\delta_{\mathrm{agg}}/dt>0$) again.}.
This indicates that bouncing occurs when more elastic energy is stored than the energy dissipated during stretching.
However, for the all-sticking cases with $R_{\mathrm{agg}}/r_1=50$ and $\phi=0.4$, the dissipation ratio shows a slight decrease with increasing velocity.
Thus, the ratio of dissipated energy to stored elastic energy after maximum compression is determined by the collision outcomes, making it challenging to predict the impact result based on this value.

\subsection{Summary of Energetics}
The energy conversion pathways and efficiencies discussed in this section are summarized in Figure \ref{fig:col_model}. 
During the compression phase, approximately 90\% of the kinetic energy is dissipated mainly through sliding motion, and the remaining energy is stored as elastic energy. 
The aggregate binding energy associated with the contact is very small. 
After maximum compression, the stored elastic energy is converted into aggregate kinetic energy. 
During the stretching phase, more than 70\% of this converted kinetic energy is dissipated primarily through rolling motion. 
These energy dissipations occur in a spherical region surrounding the contact surface (Figure~\ref{fig:edis_cross}).
The remaining kinetic energy is no longer even a few percent of the initial kinetic energy, but if the aggregate binding can be broken by this energy, bouncing occurs.

As mentioned in the discussion above, the dominant energy dissipation processes during compressive and stretching motions upon collision are qualitatively similar to those in static compression \citep{Tatsuuma+23} and static stretching \citep{Tatsuuma+19}, respectively.
For instance, the static compression simulations to determine how much elastic energy is stored or the static stretching simulations to determine how much energy is needed to separate aggregates can give us hints to understand the origin of the bouncing/sticking threshold.

Moreover, as shown in Figure~\ref{fig:Energy_all} (d), the aggregate binding energy $U_{\mathrm{bind}}$ is negligibly small compared to the initial kinetic energy. 
This indicates that the stuck aggregates are bound very loosely and therefore may separate when they collide with another aggregate. 
The outcome of sequential aggregate collisions should be studied in future work.

\section{Discussion} \label{sec:discussion}

In this section, we first discuss one possible interpretation of the strong dependence of the threshold mass required for bouncing on the filling factor (Section \ref{sec:Pcomp}). 
We then compare our results with previous work in Section \ref{sec:compare} and present an application of our findings to dust growth in protoplanetary disks in Section \ref{sec:application}. 
Finally, we discuss the limitations of our work in Section \ref{sec:limit}.

\subsection{Compressive Strength and Bouncing Conditions} \label{sec:Pcomp}

As mentioned in Section~\ref{sec:result_params}, the threshold mass for bouncing depends strongly on volume filling factor. 
In this subsection, we propose a possible origin of this strong dependence.

\begin{figure}[t]
    \centering
    \includegraphics[width=0.45\textwidth]{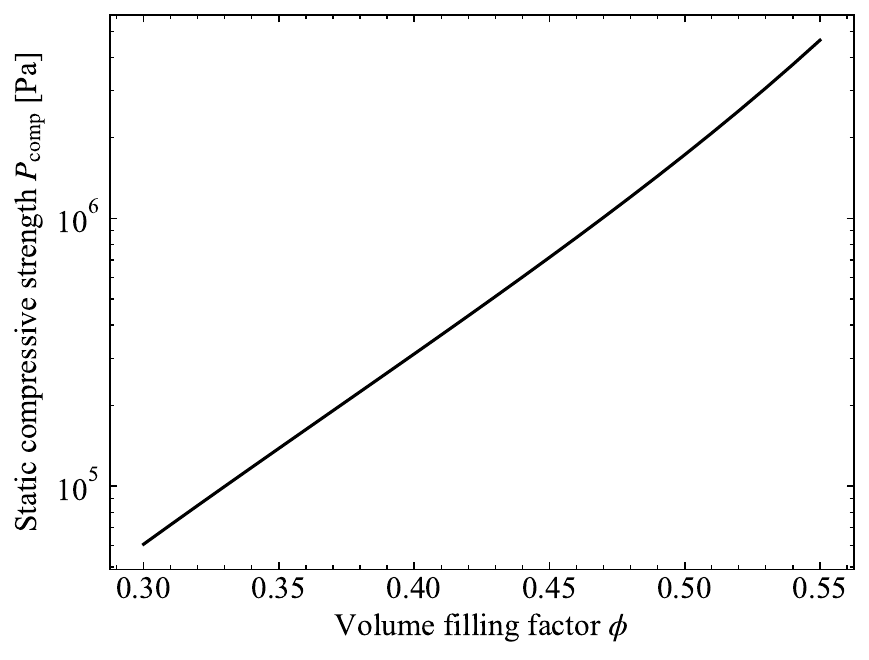}
    \caption{Compressive strength of dust aggregates consisting of ice grains with a radius of $0.1~\mathrm{\mu m}$ given by Equation~(\ref{eq:Tatsuuma_P}) \citep{Tatsuuma+23}.}
    \label{fig:P_comp}
\end{figure}

The compressive strength of dust aggregates is also sensitive to the volume filling factor for compact dust aggregates \citep{Tatsuuma+23}. 
The compressive strength is given by 
\begin{equation} \label{eq:Tatsuuma_P}
    P_{\mathrm{comp}}(\phi)=\frac{E_{\mathrm{roll}}}{r_1^3}\left(\frac{1}{\phi}-\frac{1}{\phi_{\mathrm{max}}}\right)^{-3},
\end{equation}
where $\phi_{\mathrm{max}}=\sqrt{2}\pi/6$ is the volume filling factor of the closest-packing aggregates.
As shown in Figure~\ref{fig:P_comp}, the compressive strength increases exponentially for the volume filling factor within the simulated range, $0.4\le\phi\le0.5$. 
The compressive strength increases by almost an order of magnitude {when $\phi$ is increased by 0.1}.

\begin{figure}[t]
    \centering
    \includegraphics[width=0.45\textwidth]{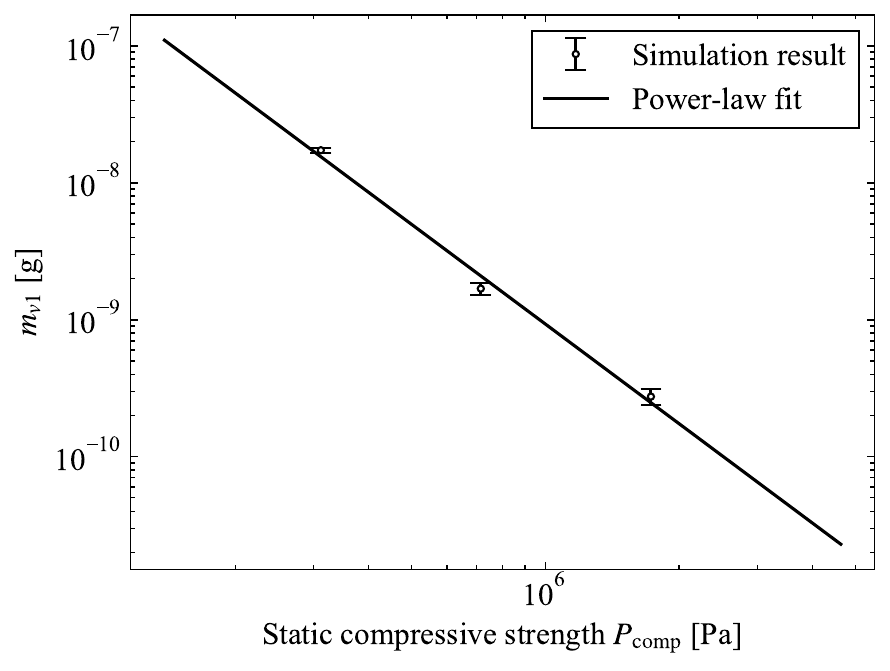}
    \caption{Same as Figure \ref{fig:result_phimv1}, but as a function of the static compressive strength $P_{\rm comp}$, using the relation between $P_{\rm comp}$ and $\phi$ shown in Figure~\eqref{fig:P_comp}. The solid line represents the power-law fit given by Equation~(\ref{eq:P_fit}).}
    \label{fig:result_Pmv1}
\end{figure}

Figure~\ref{fig:result_Pmv1} shows the threshold masses for bouncing $m_{v1}$ versus the compressive strengths $P_{\mathrm{comp}}$ obtained from simulations with three values of $\phi$. The data can be fitted by a power law
\begin{equation} \label{eq:P_fit}
    m_{v1}=2.4\times10^{-7}\left(\frac{P_{\mathrm{comp}}(\phi)}{10^{5}~\mathrm{Pa}}\right)^{-2.4}~\mathrm{g}.
\end{equation}
This expression suggests that the strong dependence of $m_{v1}$ on the filling factor can be largely attributed to the strong $\phi$ dependence of $P_{\rm comp}$.

\subsection{Comparison with Previous Work} \label{sec:compare}

Previous simulations suggest that the average coordination number is a key parameter for understanding bouncing.
\citet{Wada+11} showed that bouncing can occur when the average coordination number exceeds 6.
However, as shown in Figure~\ref{fig:nc}, the average coordination number of the compressed BCCA spheres used in our study is approximately 4.
Similarly, the average coordination number of the aggregates used in \citet{Kothe+13} is also about 4.
Experiments with larger aggregates consisting of $\sim 10^9$ monomers have confirmed bouncing even at a volume filling factor of 0.15 \citep{Langkowski+08}.
Additionally, recent experiments demonstrated that cm-sized aggregates with a filling factor of 0.1 also exhibit bouncing \citep{Schrapler+22}.
These findings suggest that the threshold average coordination number decreases as aggregate size or the number of constituent particles increases.

The impact velocity dependence of bouncing has not been well understood in previous simulations.
\citet{Wada+11} showed that bouncing occurs at intermediate impact velocities of $1~\mathrm{m\,s^{-1}} \le v_{\mathrm{imp}} \le 10~\mathrm{m\,s^{-1}}$.
Similarly, \citet{Arakawa+23} reported no significant velocity dependence in collision outcomes within this velocity range.
In contrast, our results indicate that collision outcomes do depend on impact velocity, even within the intermediate range, as described in Section~\ref{sec:result}.
Further investigations using simulations that cover a broader range of impact velocities are necessary to fully reconcile these findings with experimental results.

Additional collision simulations with varying material parameters are needed to compare our results more comprehensively with previous experiments.
By substituting $m_{v1}=1.0\times10^{-7}~\mathrm{g}$, which corresponds to the experimental result of \citet{Kothe+13}, we estimate the compressive strength of aggregates to be approximately $1.0 \times 10^5~\mathrm{Pa}$.
This value is roughly an order of magnitude higher than the compressive strength reported by \citet{Guttler+09}.
A direct comparison requires determining the scaling law of $m_{v1}$ with respect to monomer parameters, such as monomer radius and surface energy.

\subsection{Application to Protoplanetary Disks} \label{sec:application}

In this subsection, we use the results of our collision simulations to discuss the impact of the bouncing barrier on dust growth in protoplanetary disks.
Recent submillimeter polarimetric observations of protoplanetary disks have revealed that some disks produce uniformly polarized dust thermal emission \citep[e.g.,][]{Stephens+17,Stephens+23,Hull+18}, indicating that the dust aggregates responsible for the polarized emission have a maximum size of $\sim 0.1$--1 mm \citep[e.g.,][]{Kataoka+15,Yang+16,Ueda+21} and filling factors of $ \gtrsim 0.1$ \citep{Tazaki+19,Zhang+23,Ueda+24}.
The disk around IM Lup is one of the most well-studied disks producing uniformly polarized submillimeter emission \citep{Hull+18}. \citet{Ueda+24} show that IM Lup disk's bright millimeter emission and high millimeter polarization degrees require aggregates with $\phi \sim 0.3$ and a low threshold sticking velocity of $ \approx 0.3$--$1~\rm m~s^{-1}$, resulting in maximum aggregate sizes of 0.1--1 mm in the observed disk region of $\gtrsim 10~\rm au$ from the central star.

\begin{figure}[t]
    \centering 
    \includegraphics[width=\linewidth]{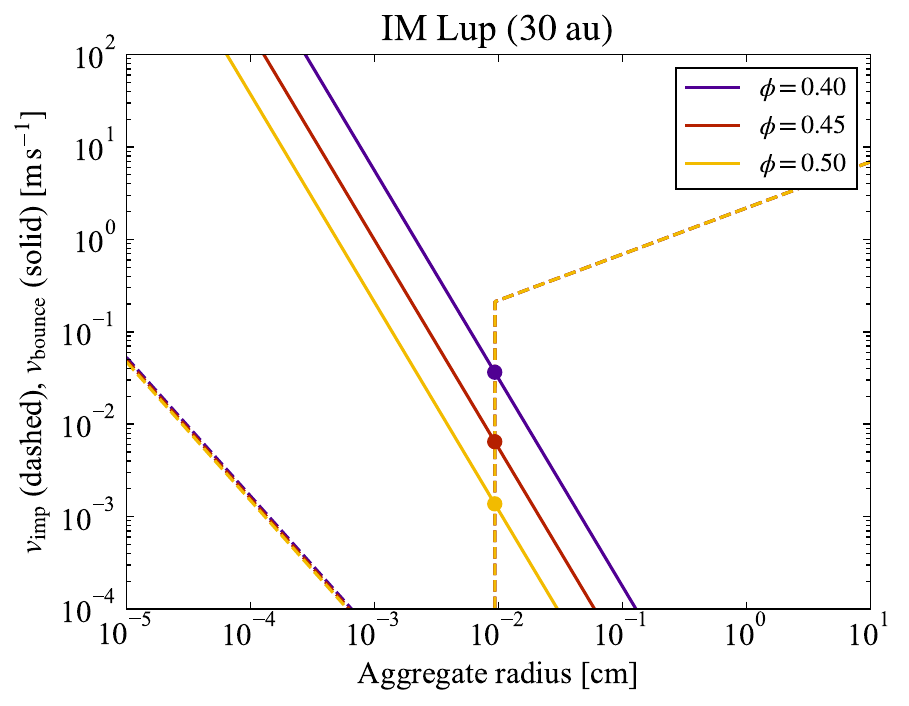} 
    \caption{Collision velocity $v_{\rm imp}$ of equal-sized dust aggregates as a function of aggregate radius at 30~au from the central star in the IM Lup disk (dashed line), compared with the bouncing threshold $v_{\rm bounce}$ given by Equation~(\ref{eq:v_bounce}) (solid lines) The circle symbols indicate the maximum aggregate size limited by bouncing collisions.} 
    \label{fig:IMLup}
\end{figure}

Here, we examine whether the bouncing barrier can explain the maximum aggregate sizes inferred for the disk around IM Lup.
We assume that aggregates grow through mutual sticking until their collision velocity $v_{\rm imp}$ exceeds the bouncing threshold $v_{\rm bounce}$ given by Equation~\eqref{eq:v_bounce}.
Following the simulation results by \citet{Dominik&Dullumond24}, which show that bouncing leads to a narrow aggregate size distribution, we only consider collisions of equal-sized aggregates.
The collision velocity we calculate includes Brownian motion and turbulence-driven relative velocity \citep{Ormel&Cuzzi07}, with the latter being dominant. Radial and azimuthal drift, as well as vertical settling, give no contribution to the relative velocity of equal-sized aggregates.
The collision velocity as a function of aggregate size $R_{\rm agg}$ depends on turbulence strength, gas temperature, and gas surface density, which we take from the IM Lup disk model of \citet{Ueda+24} at an orbital distance of 30 au. 

Figure~\ref{fig:IMLup} compare the collision velocities and bouncing thresholds for $\phi = 0.4$--0.5.
The collision velocity for the equal-sized aggregates has a cut-off at $R_{\rm agg} \approx 0.1~\rm mm$, below which the aggregates are strongly coupled to the turbulent eddies of all scales, resulting in vanishing turbulence-induced collision velocity \citep{Ormel&Cuzzi07}. 
Because of this cut-off, the maximum aggregate size given by the condition $v_{\rm imp} = v_{\rm bounce}$ converges to $100~\mathrm{\mu m}$ for $\phi \gtrsim 0.4$ \citep{Dominik&Dullumond24}. 
The predicted maximum aggregate size, indicated by the circles symbols in Figure~\ref{fig:IMLup}, is consistent with those inferred from the millimeter observations of the IM Lup disk.
For $\phi \geqslant 0.4$, the maximum aggregate size is reached when $v_{\rm imp} \approx 0.04~\rm m~s^{-1}$. 
This predicted maximum collision velocity is lower than the maximum collision velocity for sticking of 0.3--1 $\rm m~s^{-1}$ inferred by \citet{Ueda+24}. However, extrapolating Equation~\eqref{eq:v_bounce} toward lower $\phi$ suggests that the bouncing threshold would be more consistent with the inferred maximum collision velocities if $\phi \approx 0.35$.

Although the above comparison illustrates the potential of the bouncing barrier in explaining the maximum aggregate sizes in real protoplanetary disks, the bouncing barrier does not necessarily explain all observational constraints on the aggregate size distribution.
As illustrated in Figure~\ref{fig:snapshot}, bouncing collisions produce fewer small dust grains compared to fragmentation collisions \citep[see][for the mass distribution of small fragments produced by catastrophically fragmenting collisions]{Hasegawa+23}.
In fact, however, infrared observations show that protoplanetary disks typically retain a considerable amount of micron-sized grains \citep[e.g.,][]{Ueda+24}, apparently contradicting the prediction from the bouncing barrier \citep[see also][]{Dominik&Dullumond24}. 
Understanding the production rate of small dust grains in bouncing collisions and investigating its dependence on material parameters is a key topic for future research.

\subsection{Limitations of This Work} \label{sec:limit}

While we have performed simulations to investigate the bouncing barrier, additional simulations with a wider range of parameters are necessary to fully understand the conditions for bouncing.

First, there are several potentially important differences between the dust grains assumed in our simulations and those used in  previous bouncing experiments \citep[e.g.,][]{Kothe+13}. These include the monomer size (and its distribution), monomer shape, compositions, impact parameters, and aggregate shape.
However, despite these differences, our simulations qualitatively reproduce the power-law scaling between the  threshold mass for bouncing and the impact velocity, $m_{\rm crit} \propto v_{\rm imp}^{-4/3}$ (Equation~\eqref{eq:fitting_phimv1}), observed in experiments.
This suggests that the power-law exponent of $-4/3$ is robust and is insensitive to the aforementioned parameters. Future simulations should explore how the proportionality constant $m_{v1}$ in Equation~\eqref{eq:fitting_phimv1} depends on the parameters.

The aggregate temperature also influences the collision outcomes.
Recent experiments \citep{Gundlach+18,Musiolik&Wurm19} indicate that the surface energy of low-temperature ice is lower than the value assumed in our simulations.
Additionally, sintered aggregates can exhibit bouncing even at high porosity \citep{Sirono+17}.
Future work will focus on investigating the dependence of collision outcomes on surface energy and sintering effects.

\section{Summary and Conclusion} \label{sec:conclusion}

We performed three-dimensional $N$-body collision simulations of compressed dust aggregates, varying the impact velocities, aggregate sizes, and aggregate filling factors to investigate the conditions under which bouncing occurs and its parameter dependence.
The coordination numbers of compressed BCCA spheres with filling factors of 0.4, 0.45, and 0.5, which we used as the initial aggregates, were approximately 4  (Figure~\ref{fig:nc}), lower than the critical value suggested by \citet{Wada+11}.

As shown in Figure~\ref{fig:result_mv}, our collision simulations results demonstrate that the threshold aggregate mass for bouncing increases with decreasing in the impact velocity for lower than $v_\mathrm{imp} < 10~ \mathrm{m\,s^{-1}}$, and scales with impact velocity as $m_{\mathrm{bounce}}\propto v_{\mathrm{imp}}^{-4/3}$, consistent with previous experiments \citep{Kothe+13}.
Additionally, we find that the threshold aggregate mass for bouncing $m_{v1}$ decreases sharply as the aggregate filling factor increases (Figure~\ref{fig:result_phimv1}). 
We propose that the origin of this strong dependence is from the compressive strength of compressed aggregates \citep{Tatsuuma+23} (Section~\ref{sec:Pcomp}).

In Section~\ref{sec:energy}, we analyzed the energy conversion processes in selected simulations, as summarized in Figure~\ref{fig:col_model}.
Approximately 90\% of the initial kinetic energy is dissipated during compression, while the remaining 10\% is stored as elastic energy in the aggregates (region (i) in panel (b) of Figure~\ref{fig:Energy_all}).
This elastic energy is efficiently converted back into kinetic energy (middle panels of Figure~\ref{fig:energy_eff}) but over 70\% of {stored elastic energy} is dissipated during stretching, and when the remaining energy is sufficient to break the contact surfaces, bouncing occurs.

Our collision simulations suggest that dust aggregates with $\phi \gtrsim 0.4$ stop growing around $100~\mathrm{\mu m}$ (Figure~\ref{fig:IMLup}), which is consistent with the size and filling factor inferred from millimeter-wave observations \citep[e.g.,][]{Ueda+24}
However, as illustrated in Figure~\ref{fig:snapshot}, bouncing aggregates produce few small fragments, which could affect the observational appearance of protoplanetary disks.
To better understand the impact of the bouncing barrier, it will be necessary to quantify the production of small fragments and perform dust growth simulations that account for the bouncing barrier, including its fragmentary outcomes.


\vspace{\baselineskip}

The authors thank Sota Arakawa, Yukihiko Hasegawa, Yuki Yoshida and Cornelis P. Dullemond for fruitful discussions.
This work was supported by JSPS KAKENHI Grant Numbers JP22KJ1292, JP20H00182, and JP20H00205.

%

\vspace{5mm}







\bibliography{reference}{}
\bibliographystyle{aasjournal}



\end{document}